\documentclass{article}
\usepackage{spconf,amsmath,graphicx}
\usepackage{enumitem}
\usepackage{pifont}
\usepackage{multirow}
\usepackage{booktabs}
\usepackage{threeparttable}
\newcommand \footnoteONLYtext[1]
{
	\let \mybackup \thefootnote
	\let \thefootnote \relax
	\footnotetext{#1}
	\let \thefootnote \mybackup
	\let \mybackup \imareallyundefinedcommand
}

\title{audio-free prompt tuning for Language-Audio Models}
\name{Yiming Li$^{1,2}$, 
      Xiangdong Wang$^{1,\star}$,
      Hong Liu$^{1}$
      }
\address{$^1$ Beijing Key Laboratory of Mobile Computing and Pervasive Device,\\
Institute of Computing Technology, Chinese Academy of Sciences, Beijing, China. \\     
$^2$ University of Chinese Academy of Sciences, Beijing, China.  }
\begin{document}
%
\maketitle
\begin{abstract}
Contrastive Language-Audio Pretraining (CLAP) is pre-trained to associate audio features with human language, making it a natural zero-shot classifier to recognize unseen sound categories. To adapt CLAP to downstream tasks, prior works inevitably require labeled domain audios, which limits their scalability under data scarcity and deprives them of the capability to detect novel classes as the original CLAP. In this work, by leveraging the modality alignment in CLAP, we propose an efficient audio-free prompt tuning scheme aimed at optimizing a few prompt tokens from texts instead of audios, which regularizes the model space to avoid overfitting the seen classes as well. Based on this, a multi-grained prompt design is further explored to fuse global and local information. Experiments on several tasks demonstrate that our approach can boost the CLAP and outperform other training methods on model performance and training efficiency. While conducting zero-shot inference on unseen categories, it still shows better transferability than the vanilla CLAP. Moreover, our method is flexible enough even if only knowing the downstream class names. The code will be released soon. 

\end{abstract}
\section{Introduction}
\label{sec:intro}
\footnoteONLYtext{$^\star \,\, \text{Corresponding author.}$}
Audio classification, which aims at mapping an audio clip into one or more sound classes, can assist in perceiving physical environments. Recently, many approaches have achieved notable success on audio classification tasks with the emergence of large-scale audio datasets \cite{gemmeke2017audioset}, powerful network architectures \cite{chen2022hts}, and effective learning theories \cite{xiao2023semanticac}. However, they generally require a number of downstream audios for training and cannot be applied to recognize unseen categories out of the training set once tuned. With the supervision of natural language, Contrastive Language Audio Pre-training (CLAP) \cite{elizalde2023msclap, wu2023laionclap, wu2022wav2clip, guzhov2022audioclip, manco2022musicclip, huang2022mulan} resorts to a novel paradigm to tackle the above problems with two modality encoders learning a joint space for audio and text embeddings, which empowers it with zero-shot classification ability. Although CLAP provides a well-structured feature space for downstream training and demonstrates promising transferability, its zero-shot performance is still far away from real-world applications. As a result, how to efficiently enhance the classification ability of CLAP on targeted classes while reserving its generalization ability on unseen classes remains a tricky problem. 

To further adapt CLAP on downstream classification tasks, conventional Finetune updates the whole CLAP audio branch, which generally demands a substantial amount of labeled data to reach promising results. Linear Probe suggested by CLIP \cite{radford2021clip}, which solely tunes the downstream classifier, is proved to be more efficient but still ill-suited for few-shot settings. In pursuit of better performance under data scarcity, methods such as Treff adapter \cite{liang23treff} and TIP adapter \cite{zhang2022tip} utilize a key-value cache model from the training set and update the prior knowledge in CLAP by feature retrieval, which can achieve better performance than traditional few-shot algorithms, such as ProtoNet \cite{snell2017protonet}. However, the above methods discard the CLAP text encoder while newly adding learnable weights to train on a fixed number of categories, which will deprive their power to detect unseen classes as CLAP. Prompt tuning, explored in vision-language tasks \cite{zhou2022coop, khattak2023maple}, throws light on the above issue by converting the hand-crafted prompt template (e.g., ``this is a sound of'') into learnable tokens while reserving the CLAP text encoder. Although it performs well under few-shot settings and can be used to recognize unseen sound events, the learned prompt is relatively easy to overfit the seen classes and shows performance drops when testing on other datasets. Moreover, besides weakening the recognition performance on novel classes as mentioned above, prior methods rely heavily on domain audios for training, which limits their applications since domain audios might be inaccessible sometimes. In some extreme cases where no audios are available, one cannot but resort to the unpromising zero-shot predictions of CLAP. Two questions then arise: \emph{Can we bootstrap CLAP even without any downstream audios? Will it also boost the recognition ability on unseen classes?}
 
In this paper, we argue that it is feasible to treat texts as audios for training. The intuition is that the feature generated by the audio encoder will be close to the encoded text feature of its corresponding caption, so it is reasonable to extract text features from captions as alternatives for training. Compared to audios, the texts are easy to gather, and their class labels can be directly derived. We then borrow the idea of prompt tuning but train it with collected captions, resulting in a novel audio-free prompt tuning paradigm for classification tasks. To further exploit the power of prompt, we devise a multi-grained prompt tuning scheme, which aims to combine coarse-grained and fine-grained modality information to improve performance. During training, the CLAP audio encoder is excluded, making it a self-distillation process for the CLAP text encoder to regularize the parameter space of learnable prompts, thereby enhancing its generalization ability on novel classes. We refer to the above method as Prompt Tuning from Text ({\bf PT-Text}). 

Extensive experiments with different CLAP models are conducted on single-label and multi-label classification tasks. The results show that PT-Text can promote CLAP without any domain audios and outperform prior methods using few-shot audios on training efficiency and model performance. Besides, the competence to classify unseen classes is examined by a source-to-target test, and the effects of multi-grained prompts are verified by ablation studies. Moreover, even if only knowing the target class names, PT-Text can still work with hand-crafted templates, indicating its flexibility.  
 

\begin{figure}[htp]
  \centering
  \centerline{\includegraphics[scale=0.35]{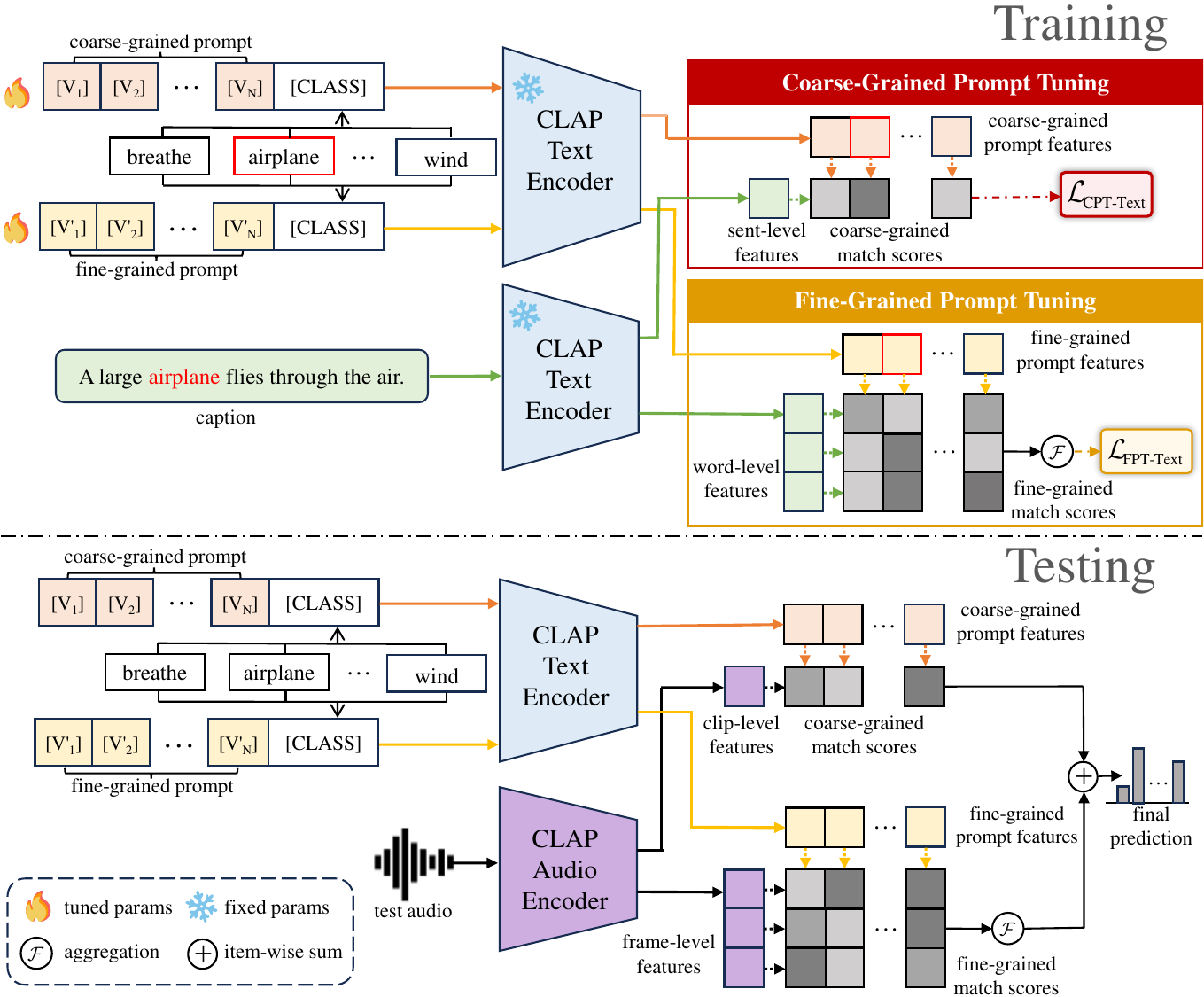}}
  \vspace{-0.75em}
  \caption{Illustration of the training and testing stage of our method.}
  \label{fig:pipeline}
\end{figure}

\section{Methodology}
\label{sec:method}
\vspace{-0.5em}
\subsection{Preliminary}
\subsubsection{Zero-shot Prediction of CLAP}
CLAP naturally fits zero-shot classification tasks. Given a test audio $x_i$, its clip-level feature $f(x_i) \in \mathbb{R}^D$ can be extracted by the CLAP audio encoder $f$, where $D$ is the dimension of feature space shared by both modalities. And the class-wise embeddings $\{w_c\}_{i=c}^{C}$ ($w_i \in \mathbb{R}^D$) can be generated by feeding prompt templates, which usually have a form of ``this is a sound of [CLASS]'', into the CLAP text encoder $g$, where $C$ denotes the number of sound classes in downstream dataset and [CLASS] can be instantiated by $C$ specific category names, such as ``bark''. Then the match score between the $i$-th sample and the $c$-th category is defined as $s(x_i, w_c) = <f(x_i), w_c>$, where $<\cdot, \cdot>$ is the cosine similarity function. For single-label tasks, the prediction is derived by $\hat{y}_i=\operatorname{argmax}_{c=1 \cdots C}<f\left(x_i\right), w_c>$. For multi-label tasks, one can compute rank-based metrics based on the above scores.
\vspace{-1.0em}
\subsubsection{Prompt Tuning with Labeled Audios}
To bootstrap CLAP, We propose to introduce learnable prompts from vision-language tasks \cite{zhou2022coop} to substitute the hand-crafted templates utilized in the original CLAP. Formally, the learnable prompt is formulated as $V=[v_1]\cdots[v_j]\cdots[v_N][\text{CLASS}]$, where $[v_j]$ denotes the $j$-th learnable token and $N$ is its length. By replacing [CLASS] with class names and forwarding $V$ to the CLAP text encoder, one can get the class-wise prompt features $\{u_c\}_{c=1}^C$ ($u_c \in \mathbb{R}^D$), which can be applied to redefine the match score as $s(x_i, c) = <f(x_i), u_c>$. We then take the single-label classification task as an example to illustrate how to tune the prompt with audios (we refer to this training scheme as \emph{PT-Audio}). Assume that we can get $K$ downstream audios and their one-hot labels $\{x_k, y_k\}_{k=1}^{K}$, the predicted probability of a sample belonging to class $l$ can be obtained by applying Softmax over the match score $s(x_k, l)$. Then, the parameters of the original CLAP are frozen while the prompt tokens can be updated by the following loss,
\vspace{-0.5em}
\begin{equation}
\label{eq_celoss}
\mathcal{L}_{\text{PT-Audio}} = -\sum_{k = 1}^{K}\sum_{l = 1}^{C} y_{k,l} \log\frac{e^{s(x_k, l)/\tau}}{\sum_{c = 1}^{C} e^{s(x_k, c)/\tau}}
\end{equation}
\vspace{-0.5em}
where $\tau$ is the logit scale in CLAP and $y_{k, l}$ is the $l$-th element of $y_k$.

\subsection{Prompt Tuning from Texts}
\subsubsection{Motivation and Overview of Our Method}
One might notice from Equation \ref{eq_celoss} that the audio feature $f(x_k)$ and its label $y_k$ are the things we actually require for PT-Audio. Since CLAP audio encoder and text encoder are aligned, if the caption of the audio (denoted as $\tilde{x}_k$, where $\tilde{x}_k$ is the matched caption of audio $x_k$) were accessible, then it would be feasible to replace $f(x_k)$ with $g(\tilde x_k)$ ($g$ is the CLAP text encoder) to approximately compute the class-wise match scores. Moreover, labeling a sound description is relatively easy. For instance, ``An airplane flies through the air'' can be classified as the sound class ``airplane'' by simple string matching. 

Therefore, we propose to collect and annotate a large set of captions from other sources and solely leverage them to implement PT-Text. As shown in Figure \ref{fig:pipeline}, two sorts of prompts, namely coarse-grained prompt $V$ and fine-grained prompt $V'$, are introduced to gather multi-grained information. While training, the frozen CLAP Text encoder is copied twice and used to encode sentence- and word-level caption features, as well as the class-wise coarse- and fine-grained prompt features respectively. Then, the annotated caption labels are applied to supervise the match scores acquired by calculating cosine similarity between caption features and prompt features of the same granularity. During testing, the CLAP audio encoder extracts clip- and frame-level features of test audios before computing two match scores with class-specific prompt features produced by the coarse- and fine-grained prompt via cosine similarity. Finally, the two scores are added together to reach ensemble predictions.
\vspace{-0.75em}
\subsubsection{Preparation of Captions}
In order to collect plentiful and informative captions for each class $c$ in a downstream task, we devise the following strategy:
\begin{itemize}[labelsep = .5em, leftmargin = 0pt, itemindent = 0.5em]
\item {\bf Text Source} \quad To ensure reproducibility, we solely use captions from public audio caption datasets, namely WavCaps \cite{mei2023wavcaps}, AudioCaps \cite{kim2019audiocaps} and Clotho \cite{drossos2020clotho}, as the textual data source. Note that crawling texts from the web or generating them with Large Language Models \cite{brown2020gpt} are also workable without regard to reproducibility.
\item {\bf Collecting Pipeline} \quad At first, a synonym dict is built by involving synonyms of each class name in the downstream dataset. Then, we maintain a collection set for each class with captions containing at least one item in its synonym dict. Next, we recheck each class's  collection set and move out captions involving items from synonym dict of other classes (for multi-label tasks, we skip this step). Finally, we balance the cardinal of these sets so that each of them has approximately $L$ captions. For sets much smaller than $L$, we manually design some templates, such as ``[CLASS] sound in the background'', and replace [CLASS] with specific class names.
\end{itemize}

Although the above strategies may introduce much noise due to the complexity and diversity of language descriptions, they can be done quickly and lead to acceptable results, as will be shown.
\vspace{-0.75em}
\subsubsection{Multi-grained Prompt Training}
In this section, we will elaborate on how to train multi-grained prompts with only supervision from texts. Denote the labeled caption collections as $\{t_m, z_m\}_{m=1}^{M}$ where $z_m$ is the according label (one-hot label for single-label tasks while multi-hot label for multi-label tasks) of caption $t_m$. As for \emph{coarse-grained prompt tuning} (CPT), we optimize the following $\mathcal{L}_{\text{CPT-Text}}$ loss and Equation \ref{eq_text_celoss_sl}, \ref{eq_text_celoss_ml} are its single-label and multi-label versions respectively,
\vspace{-0.5em}
\begin{equation}
\label{eq_text_celoss_sl}
\mathcal{L}_{\text{CPT-Text}} = -\sum_{m = 1}^{M}\sum_{l = 1}^{C} z_{m,l} \log\frac{q(t_m, u_l)/\tau}{\sum_{c = 1}^{C} q(t_m, u_c)/\tau}
\end{equation}
\vspace{-1.5em}
\begin{equation}
\label{eq_text_celoss_ml}
\mathcal{L}_{\text{CPT-Text}} = \sum_{m = 1}^{M}\sum_{i\in z_m^{+}}\sum_{j\in z_m^{-}}\max(0, 1-q(t_m, u_i)+q(t_m, u_j))
\end{equation}
where $q(t_m, u_c) = <g(t_m), u_c>$ is the match score between sentence-level feature $g(t_m)$ and coarse-grained prompt feature $u_c$ for class $c$, $z_m^{+}=\{l|z_{m,l}=1\}$, $z_m^{-}=\{l|z_{m,l}=0\}$ and Equation \ref{eq_text_celoss_ml} is a typical ranking loss in multi-label learning \cite{gong2013rankloss}. 

However, some local patterns can also provide informative cues, especially when the global features are overwhelmed by prominent sound events in multi-label clips. Therefore, we introduce the fine-grained prompt $V'$ and acquire its class-wise prompt features $\{u'_c\}_{c=1}^{C}$ accordingly. Consequently, the fine-grained match score $q'(t_m, u'_l)$ is obtained by aggregating word-level similarity scores $p_{o, l}^m$ between the $o$-th word feature $g_o'(t_m)$ and $u'_l$,
\vspace{-0.5em}
\begin{equation}
\label{eq_p}
p_{o, l}^m = <g_o'(t_m), u'_l>
\end{equation}
\vspace{-1.5em}
\begin{equation}
\label{eq_aggre}
q'(t_m, u'_l) = \sum_{o=1}^{O} \frac{e^{p_{o, l}^m/\tau_s}}{\sum_{o = 1}^{O} e^{p_{o, l}^m/\tau_s}} p_{o, l}^m
\end{equation}
where $O$ is the number of words in caption $t_m$ and $\tau_s=0.10$ is a scaling factor. By replacing $q$ in Equation \ref{eq_text_celoss_sl} and \ref{eq_text_celoss_ml} with $q'$, we can get $\mathcal{L}_{\text{FPT-Text}}$ to guide \emph{Fine-grained Prompt Tuning} (FPT). Finally, the overall training loss is defined as $\mathcal{L}_{\text{PT-Text}} = \mathcal{L}_{\text{CPT-Text}} + \mathcal{L}_{\text{FPT-Text}}$ and the illustration can be found in Figure \ref{fig:pipeline}.

Compared to PT-Audio, PT-Text is more flexible as it does not require any audios. Although texts can not be equivalent to audios due to the modality gap and label noise, it can be alleviated by involving a large corpus, which is much more accessible than audio resources. Besides, we argue that the prompt learned from texts can generalize better on unseen classes, as the training process can be viewed as self-distillation of the CLAP text encoder, which imposes implicit regularization on the model space of learnable prompt tokens. However, for PT-Audio, the prompt may be overadjusted to fit the audio modality, thereby overfitting the seen classes.

\section{Experiments}
\label{sec:exp}
\subsection{Experimental Settings}
\label{ssec:expset}
\noindent{\bf CLAP Models} \quad We adopt two state-of-the-art Language-Audio models, namely CLAP-MS \cite{elizalde2023msclap} and CLAP-LAION \cite{wu2023laionclap}, in our experiments. The former utilizes PANNs \cite{kong2020panns} as the audio encoder and BERT \cite{kenton2019bert} as the text encoder, while for the latter, the audio and text encoder are HTS-AT \cite{chen2022hts} and RoBERTa \cite{liu2019roberta}. For both language models, we insert learnable prompt tokens right after the CLS token. 
 
\noindent{\bf Downstream Datasets} \quad Four audio classification datasets, including ESC50 \cite{piczak2015esc}, US8K \cite{salamon2014us8k},  FSDKaggle2019 \cite{fonseca2019fsdkaggle} and DCASE2019 Task4 \cite{turpault2019dcase} datasets, are chosen to conduct experiments. The detailed information is presented in Table \ref{tab:table_dataset}.  
\vspace{-1.0em}
\begin{table}[th]
\begin{threeparttable}   
  \caption{Details of four audio datasets used as downstream tasks.}
  \setlength{\tabcolsep}{1.8pt}
  \label{tab:table_dataset}
  \centering
  \begin{tabular}{cccccc}
    \toprule
    \multicolumn{1}{c}{\textbf{Dataset}} &{\textbf{Multi-label}} &{\textbf{Dur}} & {\textbf{Classes}}  & {\textbf{Metric}} & {\textbf{Setup}} \\
    \midrule
    ESC50 &  & 5s & 50 & accuracy & 5 folds  \\
    US8K &  & $\leq 4\text{s}$ & 10 & accuracy & 10 folds \\
    FSD2019 & \ding{51} & 0.3-30s & 80 & mAP & train/val/test  \\
    DCASE2019 & \ding{51} & 10s & 10 & mAP & train/val/test  \\
    \bottomrule
  \end{tabular}
  \end{threeparttable}    
\end{table}
\vspace{-0.5em}

\noindent{\bf Implementation Details} \quad For the proposed method, we empirically set the number of prompt tokens $N = 16$ before initializing them with normal distribution. The number of captions collected for each class (the hyper-parameter $L$ in Section 2.2.2) is set to 16, 128, 96 and 96 for ESC50, US8K, FSD2019 and DCASE2019. More training details can be found in our released code.
\vspace{-0.5em}
\subsection{Comparison with The Original CLAP and Other Methods}
To verify the effectiveness of PT-Text, we conduct experiments on the four downstream datasets with CLAP-MS and CLAP-LAION backbone. As shown in Table \ref{tab:table_improve}, PT-Text obtains consistent improvements on all datasets over the original CLAP without using domain audios, with more significant performance improvements (nearly 0.15 absolute mAP gains) observed on the multi-label datasets.

\begin{figure}[t]
  \centering
  \centerline{\includegraphics[scale=0.5125]{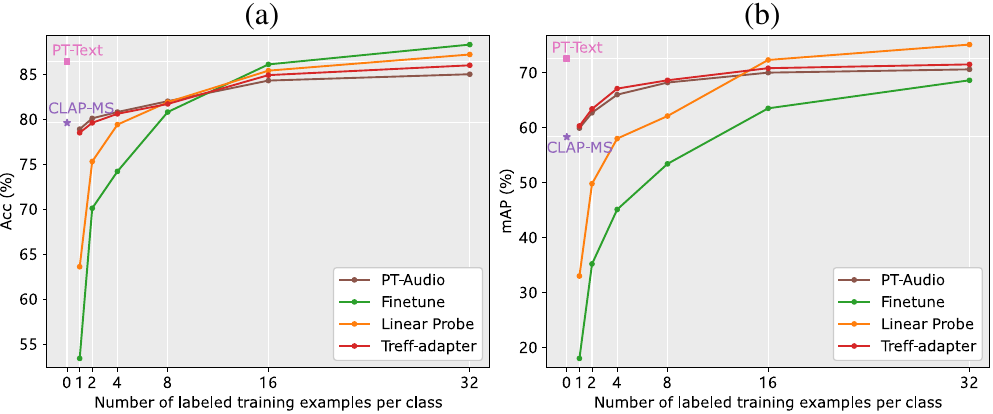}}
  \vspace{-0.75em}
  \caption{Averaged accuracy (\%) on ESC50 and US8K datasets (a) and averaged mAP (\%) on FSD2019 and DCASE2019 (b) for different training methods under few-shot settings.}
  \label{fig:fs_res}
\end{figure}

We further conduct experiments with different training methods under few-shot settings based on CLAP-MS. The results concerning evaluation metrics and training efficiency can be found in Figure \ref{fig:fs_res} and Table \ref{tab:table_param}. Generally, lightweight few-shot algorithms, such as PT-Audio and Treff adapter, perform well with fewer labeled audios. As the number of labeled examples increases, they show relatively steady performance. Nevertheless, for Finetune and Linear Probe, which generally require a lot more training resources, more labeled samples are demanded to reach promising results, especially under multi-label settings. As shown in Figure \ref{fig:fs_res} (a), when comparing PT-Text with these methods on single-label classification performance, it surpasses Finetune and Linear Probe under the 16-shot setting and also beats few-shot algorithms under the 32-shot setting. While for the multi-label classification performance shown in Figure \ref{fig:fs_res} (b), our PT-Text outperforms other methods, except Linear Probe, under the 32-shot setting. Moreover, PT-Text demonstrates superior training efficiency, with approximately 0.01M trainable parameters and 1.5 minutes of convergence time required for training.
\label{ssec:improveclap}
\begin{table}[t]
\begin{threeparttable}   
  \caption{Improvements over CLAP on four downstream datasets.}
  \setlength{\tabcolsep}{5.75pt}
  \label{tab:table_improve}
  \centering
  \begin{tabular}{ccccc}
    \toprule
    \multicolumn{1}{c}{\textbf{Methods}} &{\textbf{ESC50}} & {\textbf{US8K}} & {\textbf{FSD2019}} & {\textbf{DCASE2019}} \\
    \midrule
    CLAP-MS & 0.826 & 0.766 & 0.496 & 0.671 \\ 
    PT-Text & 0.896 & 0.830 & 0.612 & 0.834 \\
    \midrule
    CLAP-LAION & 0.905 & 0.762 & 0.561 & 0.725 \\ 
    PT-Text & 0.939 & 0.833 & 0.714 & 0.866 \\ 
    \bottomrule
  \end{tabular}
  \end{threeparttable}    
\end{table}
\vspace{-0.5em}
\begin{table}[t]
\centering
\begin{threeparttable}   
  \caption{Training efficiency of different methods averaged on ESC50 and US8K datasets, where training time is recorded on a single NVIDIA RTX 3090 GPU. Note that the 16-shot setting is adopted for training methods requiring audio labels.}
  \setlength{\tabcolsep}{10.5pt}
  \label{tab:table_param}
  \centering
  \begin{tabular}{ccc}
    \toprule
    \multicolumn{1}{c}{\textbf{Methods}} &{\textbf{\#Trainable Params}} & {\textbf{Training Time}} \\
    \midrule
    Finetune  & 80 M & 21.2 min \\
    Linear Probe   & 4.2 M & 14.9 min \\ 
    Treff Adapter  & 0.6 M & 3.21 min \\ 
    PT-Audio   & $\approx{0.01} \text{M}$ & 2.37 min \\  
    PT-Text   & $\approx{0.01} \text{M}$ & 1.43 min \\  
    \bottomrule
  \end{tabular}
  \vspace{-0.75em}
  \end{threeparttable}    
\end{table}
\begin{table}[t]
\centering
\begin{threeparttable}   
  \caption{Performance of different granular prompts (based on CLAP-MS) on four downstream datasets.}
  \setlength{\tabcolsep}{3pt}
  \label{tab:table_multi_grained}
  \centering
  \begin{tabular}{ccccccc}
    \toprule
    \multicolumn{1}{c}{\textbf{CPT}} & {\textbf{FPT}} &{\textbf{ESC50}} & {\textbf{US8K}} & {\textbf{FSD2019}} & {\textbf{DCASE2019}} & {\textbf{Average}} \\
    \midrule
    \ding{51}& & 0.886 & 0.816 & 0.577 & 0.801 & 0.770 \\
    & \ding{51} & 0.875 & 0.814 & 0.593 & 0.828 & 0.777 \\ 
    \ding{51}& \ding{51} & 0.896 & 0.830 & 0.612 & 0.834 & 0.793 \\
    \bottomrule
  \end{tabular}
  \vspace{-0.5em}
  \end{threeparttable}    
\end{table}
\begin{table}[h]
\centering
\begin{threeparttable}   
  \caption{Source-to-target generalization test trained with our method and PT-Audio. For the latter, we utilize 96 and 64 examples per class for US8K and PT-Audio as a larger one brings about trivial improvements. CLAP-MS is selected as the backbone.}
  \setlength{\tabcolsep}{7.25pt}
  \label{tab:table_cross}
  \centering
  \begin{tabular}{cccc}
    \toprule
    \multicolumn{1}{c}{}  & \multicolumn{1}{c}{\textbf{Source}} & \multicolumn{2}{c}{\textbf{Target}} \\ 
    \cmidrule(lr){2-2}\cmidrule(lr){3-4}
    {\textbf{Methods}} &\textbf{US8K} & \textbf{ESC50} & \textbf{FSD2019}\\
    \midrule
    PT-Audio (96 shots) & 0.835 & 0.794 & 0.443 \\
    PT-Text  & 0.830 & 0.861 & 0.538 \\
    \midrule
    {\textbf{Methods}} &\textbf{DCASE2019} & \textbf{ESC50} & \textbf{FSD2019} \\
    \midrule
    PT-Audio (64 shots) & 0.837 & 0.803 &  0.464 \\
    PT-Text  & 0.834 & 0.866 & 0.569 \\
    \bottomrule
  \end{tabular}
  \vspace{-0.5em}
  \end{threeparttable}    
\end{table}
\subsection{Ablation Studies}
\label{ssec:}
To thoroughly investigate the effects of each module,
we conduct a series of ablation studies on the effects of different data types used for prompt tuning, the proposed multi-grained prompt design, the generalization of learned prompt, and the number of prompt tokens.

\begin{figure}[htp]
  \centering
  \centerline{\includegraphics[scale=0.325]{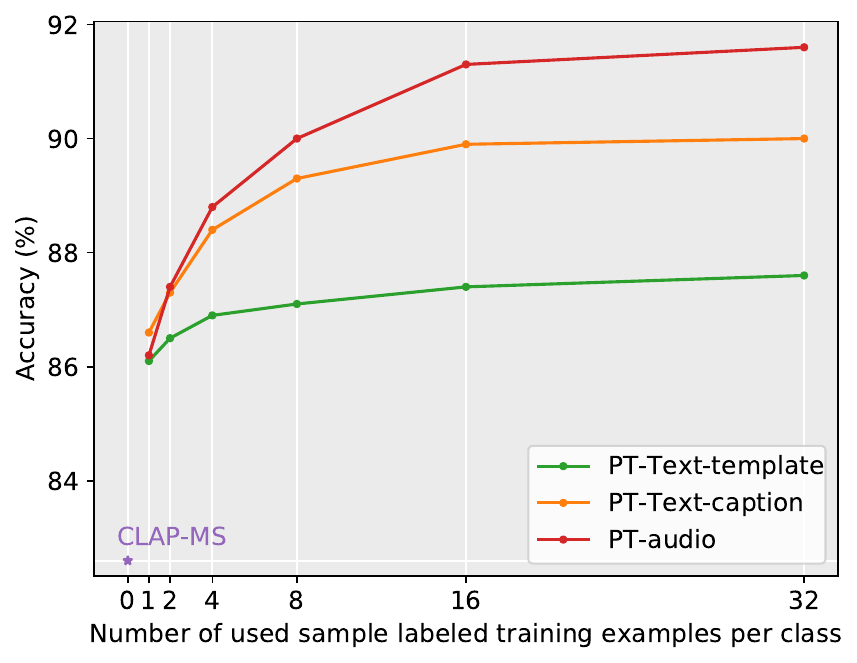}}
  \vspace{-0.75em}
  \caption{Evaluation accuracy on ESC50 trained with different numbers of training samples from different data sources.}
  \label{fig:fs_type}
\end{figure}

\noindent{\bf The effects of texts v.s. audios} \quad To examine the effects of taking captions as audios for prompt tuning, we train the model with the same number of labeled audios (denoted as PT-Audio in Figure \ref{fig:fs_type}) and compare it with the proposed PT-Text trained using collected captions (represented as PT-Text-Caption in Figure \ref{fig:fs_type}) on the ESC50 dataset. As we can see, training with 16 captions shows similar performance as training with 8 labeled audios, indicating that although applying captions is suboptimal due to the modality gap, it can still boost the classification capability with only a few captions. To further demonstrate the flexibility of PT-Text, we replace all captions with the same number of manually designed templates concatenated with downstream class names (denoted as PT-Text-template in Figure \ref{fig:fs_type} and the examples of templates can be found in Section 2.2.2), which means that the only things we need to know are class names in downstream datasets. Surprisingly, it can also bring about a 4\% absolute accuracy gain with only 4 examples per class, suggesting that we can enhance CLAP without any external knowledge.

\noindent{\bf The effects of multi-grained prompts.} \quad Table \ref{tab:table_multi_grained} presents the results of our method trained with different types of prompts. As shown, the coarse-grained prompt (CPT) is more beneficial to single-label classification tasks. By contrast, the fine-grained prompt (FPT) contributes more to multi-label tasks. We argue that for multi-label clips, the local patterns of non-significant sound classes, which are easily suppressed by some dominant sound events in coarse-grained features, can be better extracted and aggregated by the fine-grained prompt. Moreover, due to the complementary information provided for model decision, the fusion of multi-grained prompts can further improve the average performance from 0.770 and 0.777 to 0.793 for CPT and FPT, respectively. 

\noindent{\bf The generalization ability of the prompt.} \quad 
We design a source-to-target test to measure the transferability of leaned prompts, where prompts learned on source datasets are directly applied as final prompts without additional training when testing on target datasets. Table \ref{tab:table_cross} reports the results. Note that the number of classes in source datasets is much smaller than that in target datasets as seen in table \ref{tab:table_dataset}, which means that there are many unseen classes when transferring the learned prompt. It can be observed that PT-Text leads to more promising generalization ability, even surpassing the original CLAP largely on target datasets. However, similar to some insights found for prompt tuning in language-vision tasks \cite{zhou2022cocoop}, PT-Audio may overfit the audio patterns of seen classes and even show performance degradation, with an average drop of 0.03 on accuracy for ESC50 and 0.05 on mAP for FSD2019, on unseen classes when comparing with the zero-shot CLAP. Besides, it is worth mentioning that other training methods presented in Figure \ref{fig:fs_res} lose the property of open-set recognition since they leave aside the textual information.

\begin{figure}[h]
  \centering
  \centerline{\includegraphics[scale=0.325]{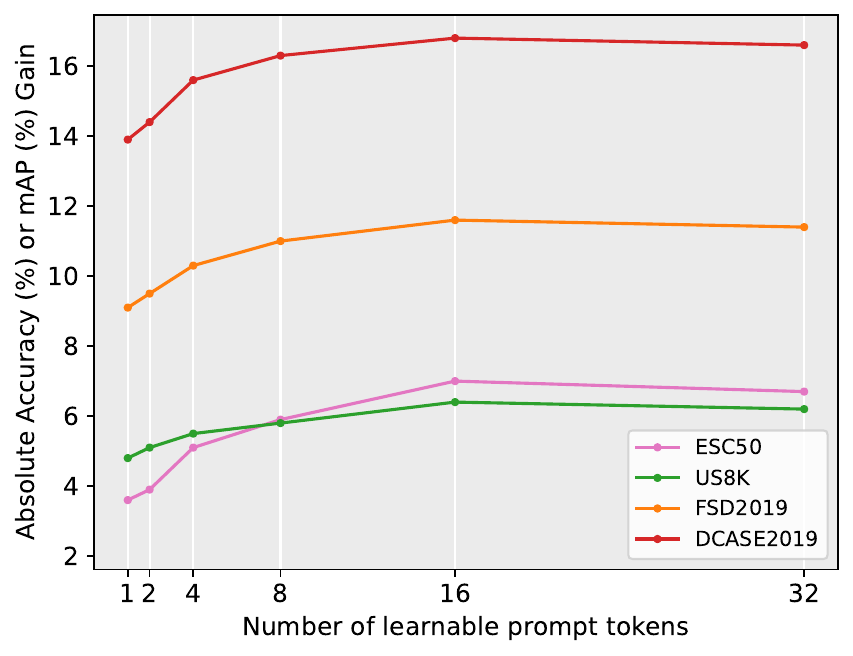}}
  \vspace{-0.75em}
  \caption{Absolute accuracy (\%) or mAP (\%) gain over original MS-CLAP when using different numbers of prompt tokens.}
  \label{fig:len_pr}
\end{figure}

\noindent{\bf The length of the learnable prompt.} \quad Figure \ref{fig:len_pr} summarizes the accuracy (or mAP for multi-label datasets) gained over original CLAP-MS using a varied number of prompt tokens on four downstream datasets. Noticeably, our method can obtain considerable improvements on all datasets even with a single prompt token. As the prompt length increases to 16, further improvements can be witnessed, with a longer prompt resulting in overfitting problems.

\vspace{-1.25em}
\vspace{-1.0em}
\section{Conclusion}
\label{sec:print}
In this paper, we develop a novel audio-free scheme to adapt Language-Audio models on downstream tasks, which optimizes a few multi-grained prompt tokens from easily accessible texts. Experiments indicate that the proposed method can boost CLAP models without labeled audios and demonstrate competitive performance and efficiency compared to existing training methods on several classification tasks under few-shot settings. Extensive ablation studies further verify its superior transferability and flexibility.
\vfill\pagebreak

\bibliographystyle{IEEEbib}
\bibliography{strings,refs}

\end{document}